\documentclass[aps,prb,reprint,superscriptaddress,longbibliography,floatfix]{revtex4-2}
\usepackage[T1]{fontenc}
\usepackage[utf8]{inputenc}
\usepackage{amsmath,amssymb}
\usepackage[version=3]{mhchem}
\usepackage{graphicx}
\usepackage{booktabs}

\usepackage{hyperref}

\begin{document}
\title{Pressure and Magnetic Field Control of Topology in Antiferromagnetic Bilayers}

\author{Narsimha Rao Elaprolu}
\affiliation{Dipartimento di Scienze Fisiche, Informatiche e Matematiche,
Università di Modena e Reggio Emilia, via Campi 213/a, I-41125 Modena, Italy}

\author{Marco Gibertini}
\affiliation{Dipartimento di Scienze Fisiche, Informatiche e Matematiche,
Università di Modena e Reggio Emilia, via Campi 213/a, I-41125 Modena, Italy}
\affiliation{Istituto Nanoscienze -- CNR, S3, via Campi 213/a, I-41125 Modena, Italy}

\begin{abstract}
Controlling band topology in magnetic semiconductors requires strategies that couple electronic structure to experimentally accessible degrees of freedom. Here we propose a cooperative approach based on pressure and magnetic reorientation in A-type antiferromagnetic bilayers. Antiparallel layer magnetizations suppress hybridization between exchange-polarized band-edge states, limiting their response to interlayer compression. A magnetic field that aligns the layer moments removes this spin mismatch, enabling pressure-enhanced hybridization to drive a band inversion. The magnetization direction then controls the symmetries of the inverted bands and whether spin--orbit coupling produces a Chern-insulating gap or symmetry-protected nodal crossings. First-principles calculations for bilayer CrSBr illustrate this mechanism within a compressed orthorhombic structural model. Out-of-plane magnetization yields a Chern insulator with a chiral edge state, whereas in-plane magnetization supports a two-dimensional Weyl semimetal with node-connecting edge states. Continuous spin canting further demonstrates how magnetic reorientation can drive the transition and stabilize the Chern phase before full magnetic saturation. Although material-specific transition conditions depend on electronic correlations and structural stability under compression, the underlying strategy identifies interlayer hybridization as a link between magnetic order and band topology. These results provide a design principle for externally tunable topological phases in layered antiferromagnetic semiconductors.
\end{abstract}

\maketitle

Chern insulators combine an insulating bulk with chiral edge channels and a quantized Hall response arising from broken time-reversal symmetry.\cite{Haldane1988} Their potential for low-dissipation electronics motivates the search for magnetic materials whose topology can be controlled externally. The quantum anomalous Hall effect has been observed in magnetically doped topological-insulator films and intrinsic magnetic MnBi$_2$Te$_4$,\cite{Chang2013,Deng2020} but experimentally established platforms remain limited.

First-principles calculations have proposed several routes to Chern insulating states, including heavy-element layers on magnetic substrates,\cite{Garrity2013} ferromagnetic transition-metal halides,\cite{Huang2017Halides,Sheng2017} and magnetic metal--organic frameworks.\cite{Ni2018} Doped CrI$_3$ also supports predicted Chern phases tunable through strain and electronic correlations.\cite{Baidya2018} High-throughput calculations have broadened the candidate space while highlighting the sensitivity of some predictions to the exchange--correlation approximation.\cite{Olsen2019} These studies motivate strategies that combine favorable electronic structure with an accessible means of switching topology.

Magnetization orientation provides one such control because spin--orbit coupling (SOC) ties the band structure to magnetic symmetry. First-principles studies predict orientation-driven transitions between trivial and Chern phases in monolayer OsO$_2$,\cite{Olsen2019} changes of Chern number in NiAsO$_3$ and PdSbO$_3$,\cite{Li2022} and switching between nodal and Chern phases in TiTe.\cite{Xu2022} Magnetic alkaline-metal monochalcogenide monolayers likewise support predicted nodal-loop, Weyl-like, and quantum anomalous Hall phases selected by spin orientation.\cite{Zhou2019} Extending this principle to layered antiferromagnets introduces an additional variable, the relative alignment of neighboring layers, which can strongly modify interlayer electronic coupling.

In an A-type antiferromagnet, each layer is ferromagnetic but adjacent layers have opposite magnetizations. Atomically thin CrI$_3$ and CrSBr illustrate the control afforded by this interlayer order.\cite{Huang2017,Telford2020} For strongly exchange-polarized band edges, antiparallel alignment suppresses resonant spin-conserving hybridization, whereas parallel alignment enables it. Magnetic control of this coupling has been demonstrated in CrSBr through excitonic spectroscopy and microscopic calculations.\cite{Wilson2021} Vertical transport measurements also reveal a strong dependence on interlayer magnetic order.\cite{Lin2024} Compression can amplify the difference by increasing orbital overlap, suggesting a route to magnetic control of pressure-induced band inversion.

\begin{figure*}[t]
    \includegraphics[width=\linewidth,height=0.68\textheight,keepaspectratio]{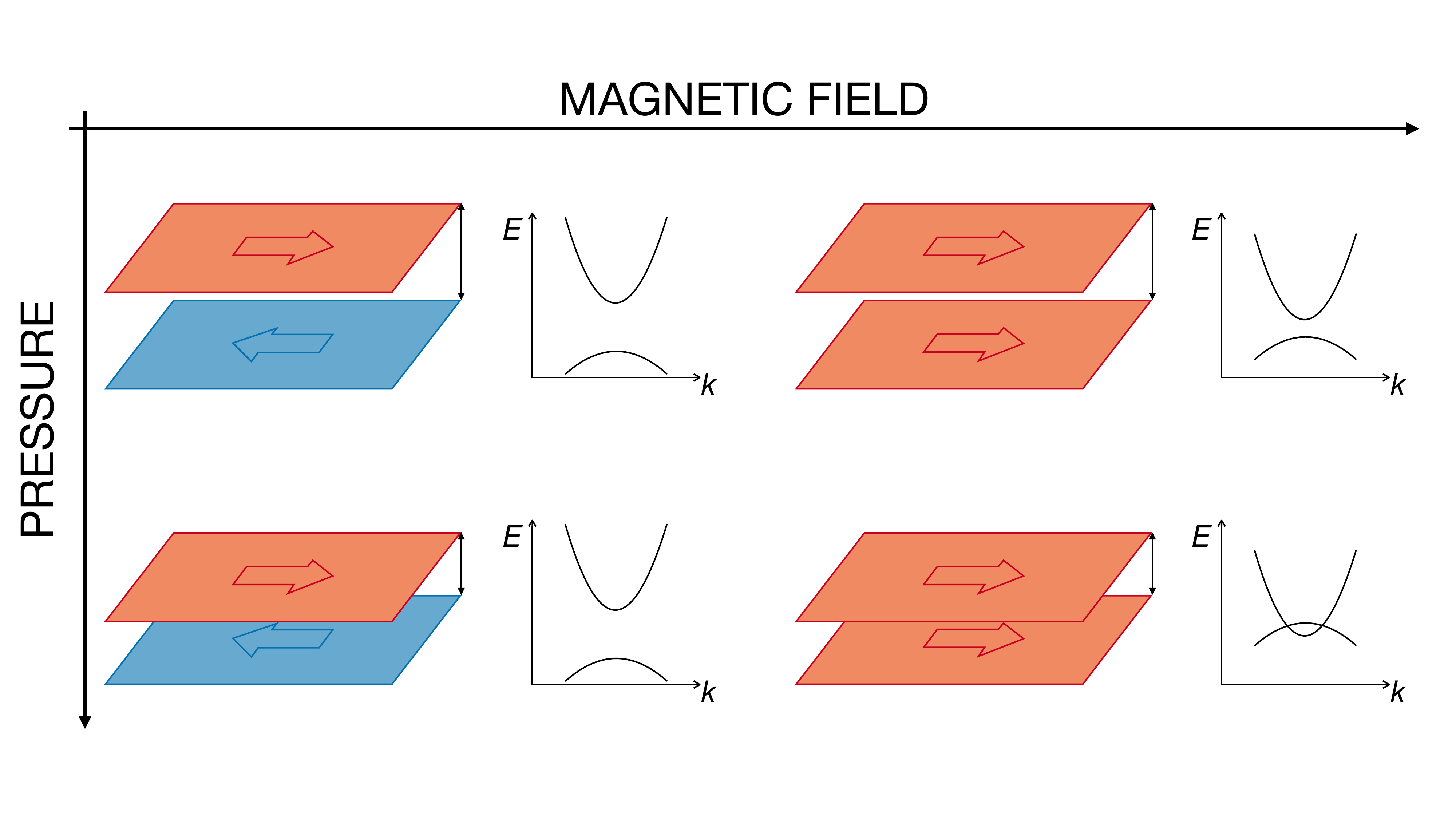}
    \caption{
    Schematic illustration of pressure- and magnetic field-induced Chern phase evolution in antiferromagnetic bilayers. An applied magnetic field favours a parallel alignment of spins in the two layers, driving a transition from the  AFM ground state to a FM configuration. Pressure reduces the interlayer separation, thus enhancing hybridization and tunnelling between the layers, especially in the FM state. The combined effect can give rise to a band inversion and thus to a topological transition from a trivial phase to a Chern insulator.   \label{fig:scheme}}
\end{figure*}

Figure~\ref{fig:scheme} summarizes this mechanism. Pressure reduces the layer separation, while a magnetic field aligns the layer moments and activates stronger hybridization. If the resulting band inversion is accompanied by an appropriate SOC gap, the bilayer can become a Chern insulator. Here we use CrSBr as a prototypical example to investigate by means of first-principles density-functional-theory (DFT) simulations this cooperative response, classify the phases selected by magnetization direction, and follow a continuous canting path from the antiferromagnetic semiconductor to the Chern phase. 

\section{Results and Discussion}
\subsection{Magnetic control of interlayer coupling}
CrSBr is an air-stable orthorhombic semiconductor with ferromagnetic Cr--S layers terminated by Br atoms and stacked along $c$. Bulk CrSBr has space group $Pmmn$ and develops A-type antiferromagnetic order near 132~K.\cite{Telford2020,Lee2021} We identify $x$, $y$, and $z$ with the crystallographic $a$, $b$, and $c$ directions, the latter being normal to the layers. The ambient-pressure magnetic easy axis lies along $b$, and applied fields can reorient the layer magnetizations.\cite{Lee2021,Wilson2021}

\begin{figure}[t]
    \centering
    \includegraphics[width=\linewidth,height=0.68\textheight,keepaspectratio]{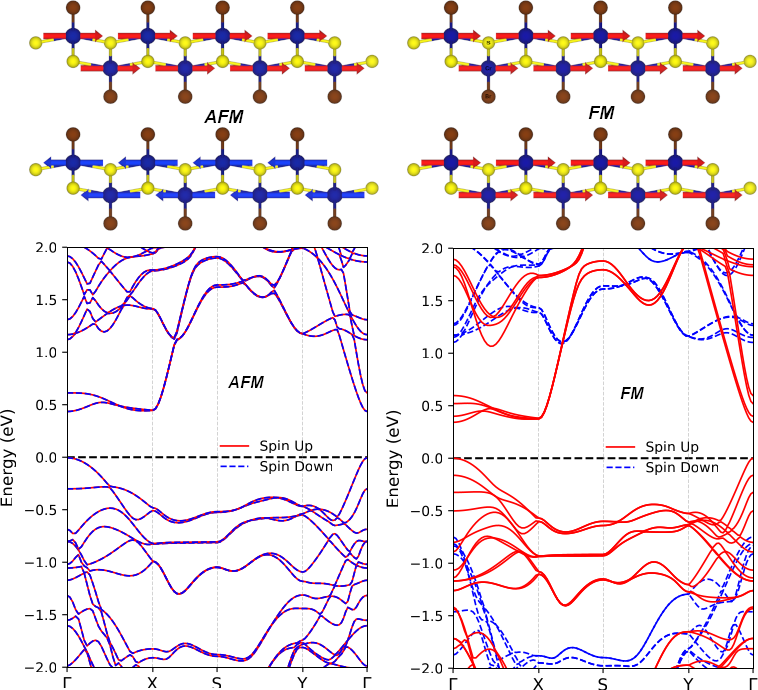}
     \caption{Top panels: Lateral views of the crystal structure of bilayer CrSBr where red and blue arrows highlight the different configurations of chromium magnetic moments in the AFM (left) and FM (right) states. Bottom panels: corresponding electronic band structures calculated using a vdW-compliant functional (vdW-DF3-opt2~\cite{Chakraborty2020}) without spin–orbit coupling effects along a high-symmetry path in the Brillouin zone. Red solid  and blue dashed lines represent spin-up and spin-down states, respectively. The zero of energy is set at the top of the valence bands,  highlighted with a horizontal dashed line.     \label{fig:crystal_and_bands}}

\end{figure}

Figure~\ref{fig:crystal_and_bands} compares antiferromagnetic (AFM) and ferromagnetic (FM) bilayers extracted from the bulk stacking. Without SOC, electronic states can be labeled by their spin projections. For the AFM bilayer spin-up and spin-down bands perfectly overlap as a result of the combined symmetry of inversion and time reversal.\cite{Lee2021} Corresponding band states on opposite layers have opposite spin character, limiting their resonant hybridization. As a consequence the bilayer energy bands retain an approximately monolayer-like dispersion.

Parallel layer moments remove the spin mismatch. The FM bands exhibit interlayer splittings, lose spin degeneracy, and have a smaller gap than the AFM bilayer. Thus magnetic alignment changes the electronic coupling between layers, providing the basis for their different responses to compression. Related calculations have shown that strain and stacking can modify magnetic order and electronic coupling in CrSBr bilayers.\cite{Hou2023}

\subsection{Pressure-driven band inversion}
\begin{figure*}
    \centering
    \includegraphics[width=0.7\linewidth]{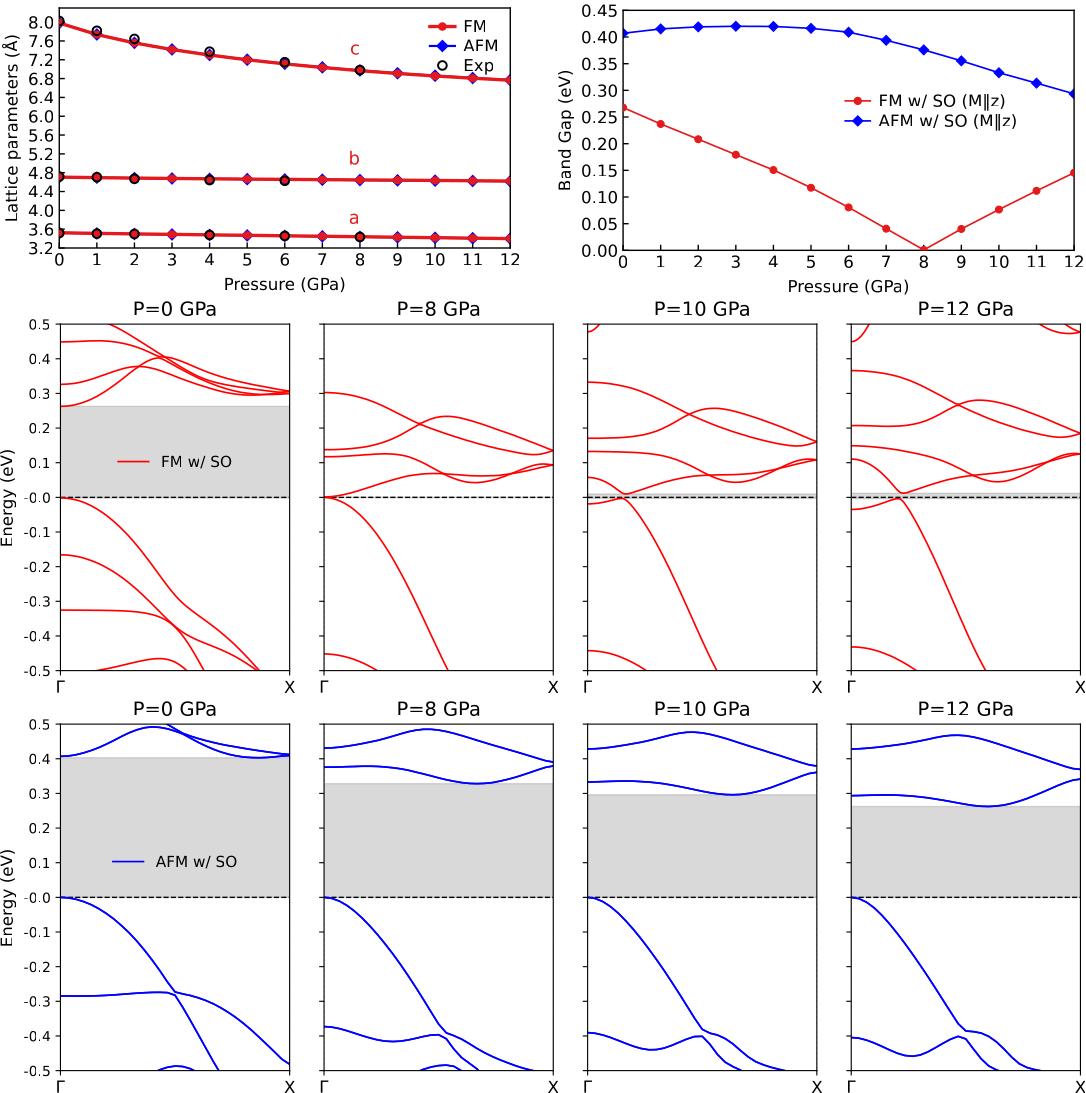}
     \caption{Effect of pressure on the structural and electronic properties of CrSBr. Top left panel: computed bulk CrSBr lattice parameters ($a$, $b$, and $c$, in \AA) as a function of applied pressure (in GPa) for the FM (red circles) and AFM (blue diamonds) state. Experimental results\cite{Langston2025} are also reported with empty circles. Top right panel: direct band separation at $\Gamma$ (in eV) of FM and AFM bilayer CrSBr as a function of pressure computed including spin–orbit coupling effects assuming a magnetization direction out of plane. Central and bottom panels: electronic band structure of FM and AFM bilayer CrSBr along the $\Gamma$-$X$ direction in reciprocal space for different values of pressure.  The dashed line marks the top of the valence bands and the gray shaded region highlights the overall band gap.    \label{fig:pressure}}
\end{figure*}

We optimize the bulk lattice parameters and atomic coordinates under hydrostatic pressure using the van der Waals (vdW) compliant functional vdW-DF3-opt2,\cite{Chakraborty2020} then extract bilayers from the relaxed structures. Pressure therefore labels the geometry inherited from compressed bulk CrSBr rather than a stress applied to a vacuum-separated slab. The upper-left panel of Figure~\ref{fig:pressure} shows that compression mainly reduces the stacking parameter $c$, while $a$ and $b$ change modestly. AFM and FM structural parameters remain close and reproduce the experimental trends plotted over the displayed range.\cite{Langston2025}

Relativistic PBEsol calculations\cite{Perdew2008} reveal sharply different electronic responses to these geometries. For moments along $z$, the direct band separation at $\Gamma$ remains finite in the AFM bilayer throughout the investigated pressure range. In the FM bilayer, it decreases rapidly and closes near 8~GPa. The selected dispersions in Figure~\ref{fig:pressure} show the approaching band edges and their inverted ordering at higher pressure. The reopened separation at $\Gamma$ exceeds the minimum insulating gap, which occurs away from $\Gamma$.

This contrast is consistent with compression-enhanced hybridization acting efficiently on parallel-spin states. The AFM spin mismatch limits the corresponding band-edge shifts, although intralayer distortions and residual coupling still change the gap. Magnetic alignment therefore converts the same structural compression into substantially different electronic responses.

The inverted states at $\Gamma$ have opposite parity in the centrosymmetric FM model. Exchanging an occupied and an unoccupied state of opposite parity changes the Chern parity of a globally gapped two-dimensional system if no additional inversions occur elsewhere.\cite{Hughes2011} This criterion motivates a topological analysis but does not establish the insulating character or determine the Chern number itself. The Chern number is a global property of the occupied Bloch subspace, compatible with inversion symmetry once time-reversal symmetry is broken.\cite{Kattan2026}

\subsection{Topology selected by magnetization direction}
\begin{figure*}
 		\centering
         \includegraphics[width=\linewidth,height=0.68\textheight,keepaspectratio]{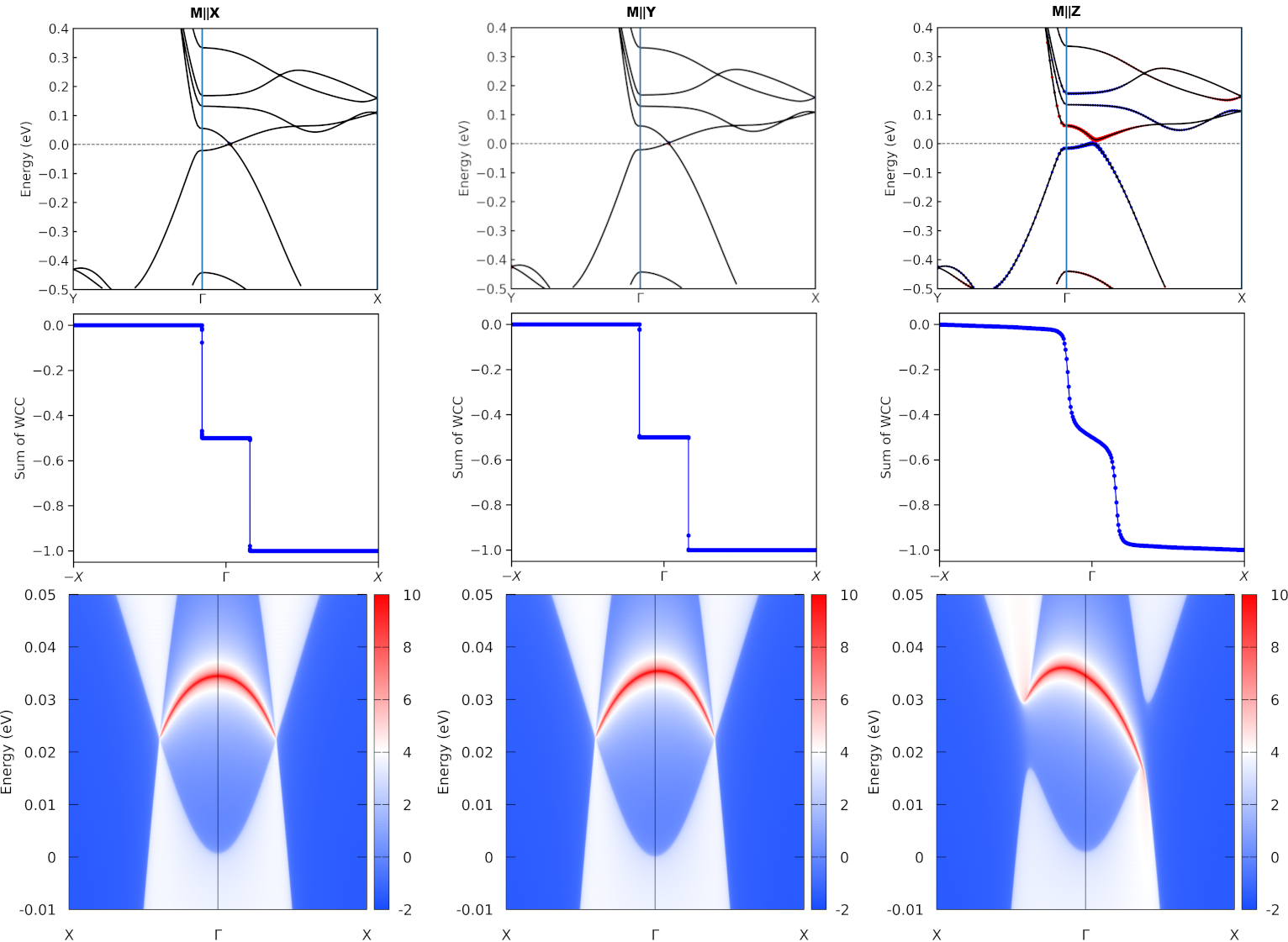}
 		 \caption{Electronic structure and topological classification of FM-CrSBr bilayer at 10~GPa for three different magnetization directions ($\mathbf{M} \parallel \mathbf{\hat x}$, $\mathbf{M} \parallel \mathbf{\hat y}$ and $\mathbf{M} \parallel \mathbf{\hat z}$, corresponding to the left, central, and right columns). Top panels: electronic band structure along the $Y$-$\Gamma$-$X$ high-symmetry path with corresponding Berry curvature projection. Red (blue) circles denote positive (negative) Berry curvature contributions, respectively, while the circle size is proportional to the magnitude of Berry curvature. Central panels: evolution of the occupied-band Zak phase computed along the $y$ (i.e. $b$) direction as a function of crystal momentum along the $x$ (i.e. $a$) direction, represented in terms of sum of Wannier charge centres (WCC). Bottom panels: momentum-resolved edge spectral function of a semi-infinite  FM-CrSBr bilayer showing the appearance of topological edge states.     \label{fig:topology}}
\end{figure*}

Figure~\ref{fig:topology} compares the compressed FM bilayer at 10~GPa for magnetization along the three principal axes. Wannier interpolation\cite{Pizzi2020,Tsirkin2021} provides the Berry curvature, the occupied hybrid-Wannier charge-center (WCC) evolution, and the edge spectra. For either in-plane orientation, the inverted bands cross along $\Gamma$--$X$. The antiunitary symmetry $C_{2z}\mathcal{T}$ leaves two-dimensional momentum invariant and reverses the out-of-plane Berry curvature, forcing it to vanish away from the nodes.\cite{Xiao2010,Xu2022}

The middle panels show the corresponding piecewise constant WCC sum for loops along $k_y$. Although the WCC is undefined for a loop passing through a node, a small residual numerical gap allows us to track sharp changes by half a lattice period, which represent $\pi$ changes of the total Zak phase across the projected nodes.\cite{Zak1989} Together with the isolated crossings, this behavior identifies a two-dimensional Weyl semimetal.\cite{Xu2022,Lu2024} These isolated crossings differ from extended nodal lines protected by mirror symmetry, as established in the three-dimensional spin--orbit metal PbTaSe$_2$.\cite{Bian2016} The edge spectral functions show branches, known as Fermi strings,\cite{Lu2024} connecting the projected nodes, consistent with the different Berry phases of the intervening momentum slices. Spin structure can influence the robustness of such edge states against vacancies, as demonstrated in models of two-dimensional Weyl semimetals.\cite{Mella2022}

Out-of-plane magnetization breaks $C_{2z}\mathcal{T}$, allowing finite Berry curvature and an SOC gap at the inverted crossings. The right column shows strong Berry curvature contributions near the avoided crossing, one negative winding of the occupied WCC sum, and a chiral edge branch traversing the gap. These complementary signatures identify a Chern insulator with Chern number $C=-1$.

This orientation-driven nodal-to-Chern conversion is related to the half Chern--Weyl mechanism proposed for TiTe.\cite{Xu2022} It also illustrates the broader role of magnetic symmetry in selecting topology, as found in OsO$_2$ and transition-metal oxide monolayers.\cite{Olsen2019,Li2022} In CrSBr, compression first brings the bands into the inverted regime, after which magnetization orientation selects the nodal or gapped phase.

\subsection{Topological switching through spin canting}
\begin{figure*}
 		\centering
              \includegraphics[width=0.7\linewidth]{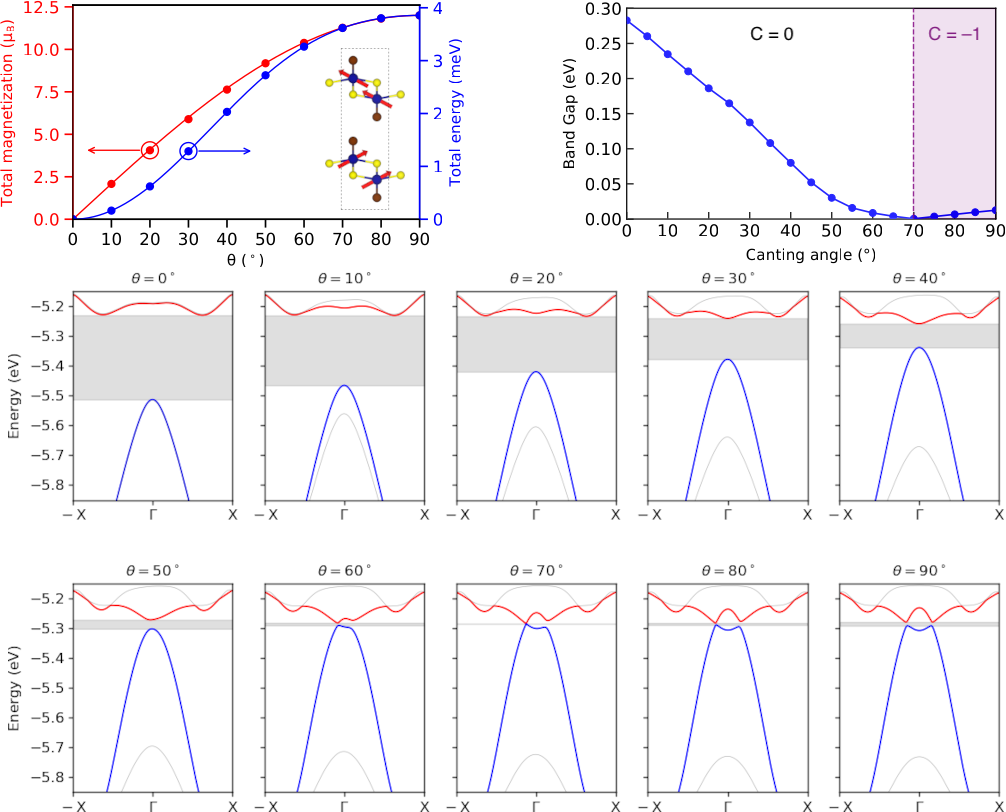}   
 		 \caption{Effect of spin canting (a proxy for magnetic-field-driven reorientation) on the electronic properties of CrSBr bilayer at 10~GPa. Top left panel: computed total energy and net magnetization as a function of the spin canting angle $\theta$. The inset shows a canted configuration where the canting angle measures the deviation of the magnetization direction from the in-plane direction, which can be obtained by applying a vertical magnetic field. An angle $\theta=0^\circ$  corresponds to an in-plane AFM configuration, while $\theta = 90^\circ$ is associated with an out-of-plane FM state. Top right panel: Energy band gap as a function of the canting angle. The gap closes at $\theta\approx70^\circ$ and then re-opens, signalling a topological phase transition where the Chern number changes from trivial ($C=0$) to non-trivial ($C=-1$). Bottom panels: computed electronic band structure of bilayer CrSBr along the $-X$-$\Gamma$-$X$ path for different spin canting angles from $0^\circ$ to $90^\circ$. The top valence and bottom conduction bands are marked in blue and red respectively, while the band gap is highlighted with a grey shaded region. A common zero of energy is set at the vacuum level far away from the bilayer.     \label{fig:canting}}
\end{figure*}

To connect the AFM ground state to the out-of-plane FM phase, we calculate a constrained noncollinear canting path at 10~GPa. As illustrated in Figure~\ref{fig:canting}, the layer moments have opposite in-plane components and a common out-of-plane component. Their angle $\theta$ from the plane ranges from the AFM configuration at $0^\circ$ to the FM configuration at $90^\circ$. For layer moments of magnitude $M_0$, the net moment follows $M_z=2M_0\sin\theta$, in agreement with the calculated trend.

The zero-field energy increases along this path, favoring the AFM endpoint. A perpendicular field lowers the energy of canted configurations through the Zeeman contribution $-B_zM_z$. The path consequently models the electronic response to field-driven reorientation, although an equilibrium switching field requires minimizing the full magnetic energy and checking competing configurations. Experiments show that pressure changes interlayer exchange and magnetic reorientation in bulk CrSBr, underscoring the importance of this magnetic-energy analysis.\cite{Telford2023,Pawbake2023}

As the spins cant toward a common direction, the band-edge spinors overlap more strongly and interlayer hybridization increases. The valence-band maximum rises, reducing the gap until it closes near $\theta\approx70^\circ$. At larger angles, a small gap reopens and the calculated Chern number changes from $0$ to $-1$. Intermediate spin textures break the layer-exchanging inversion symmetry of the FM endpoint, consistent with the unequal dispersions along the $-X$ and $+X$ branches.

The finite interval of nonzero gap and $C=-1$ shows that complete out-of-plane alignment is not required. At fixed compression, magnetic canting can therefore drive the bilayer across a topological boundary and stabilize the Chern phase before saturation.

Previous first-principles work also predicts pressure-induced metallization and a magnetic transition in CrSBr bilayers under vertical compression.\cite{Feng2025} Here we focus on the magnetization-dependent topology of the inverted bands within the retained structural branch. The quantitative pressure scale remains conditional on the structural and electronic approximations. Semilocal Kohn--Sham gaps can shift the inversion threshold, and predicted Chern phases can depend sensitively on the exchange--correlation treatment.\cite{Olsen2019} Moreover, experimental structural anomalies near 7.6--8~GPa overlap the calculated transition pressure.\cite{Langston2025,Kozlenko2026} The 10~GPa results thus describe the retained structural branch rather than an established equilibrium phase of compressed CrSBr. The small SOC gap, pressure-dependent magnetic anisotropy, and possible structural distortions must be resolved before assigning experimental operating conditions.

\section{Conclusions}
Pressure and magnetic alignment cooperate to control topology in the CrSBr bilayer model. Antiparallel moments suppress the electronic response to interlayer compression, whereas parallel alignment enables hybridization and band inversion near 8~GPa. At 10~GPa, in-plane magnetization yields a nodal semimetal and out-of-plane magnetization yields a Chern insulator with $C=-1$. A continuous canting path reaches the Chern phase near $70^\circ$, establishing a finite angular range of stability. Although the numerical pressure threshold requires further structural and electronic validation, the mechanism provides a general strategy for A-type antiferromagnets in which magnetic alignment activates compression-enhanced interlayer coupling.

\section{Methods}
First-principles calculations were performed within plane-wave pseudopotential DFT using Quantum ESPRESSO.\cite{Giannozzi2009,Giannozzi2017} Bulk CrSBr structures were optimized separately for FM and AFM order at fixed hydrostatic pressure by relaxing the lattice parameters and atomic positions. Interlayer dispersion interactions were treated with the nonlocal vdW-DF3-opt2 functional.\cite{Chakraborty2020} These calculations used ultrasoft pseudopotentials,\cite{Vanderbilt1990} kinetic-energy cutoffs of 90 and 720~Ry for the wave functions and charge density, respectively, and a $12\times9\times4$ Monkhorst--Pack mesh.\cite{Monkhorst1976} The energy,  force and stress convergence thresholds for relaxations were $6.0\times10^{-5}$~Ry, $1.0\times10^{-4}$~Ry/Bohr, and 0.1~kbar, respectively.

Bilayers were extracted from the pressure-relaxed bulk geometries. A two-dimensional Coulomb cutoff was applied along the layer-normal direction to suppress electrostatic interactions between periodically repeated slabs.\cite{IsmailBeigi2006,Sohier2017} The initial collinear band structures without SOC were calculated with vdW-DF3-opt2. Pressure-dependent relativistic bilayer calculations used PBEsol\cite{Perdew2008} and optimized norm-conserving Vanderbilt pseudopotentials\cite{Hamann2013} from the PseudoDojo library.\cite{vanSetten2018} SOC was included for the magnetization-direction comparison and the constrained noncollinear canting series.

Wannier90\cite{Pizzi2020} was used to construct a Wannier Hamiltonian from Cr $d$, S $p$, and Br $p$ projections. Berry curvature was evaluated by Wannier interpolation with WannierBerri.\cite{Tsirkin2021} The evolution of the occupied hybrid WCCs along $k_y$ as a function of $k_x$ and the semi-infinite edge spectral functions were calculated using WannierTools.\cite{Wu2018} The WCC winding was used to identify the Chern number in gapped configurations, while the edge spectra provided a complementary check of bulk--boundary correspondence.

\section{Acknowledgments}
We acknowledge financial support from the project ECOSISTER ``Ecosystem for Sustainable Transition in Emilia-Romagna'' (ECS$\_$00000033, CUP E93C22001100001), funded by the European Union – NextGenerationEU through the Italian National Recovery and Resilience Plan (PNRR), Mission 4, Component 2, Investment 1.5.
\bibliographystyle{apsrev4-2}
\bibliography{references}

\end{document}